\documentclass[conference]{IEEEtran}
\IEEEoverridecommandlockouts
\usepackage{cite}
\usepackage{amsmath,amssymb,amsfonts}
\usepackage{algorithmic}
\usepackage{graphicx}
\graphicspath{{img/}}
\usepackage{float}
\usepackage{textcomp}
\usepackage{xcolor}

\usepackage[abbreviations]{glossaries-extra}

\glssetcategoryattribute{acronym}{indexonlyfirst}{true}

\newabbreviation{stft}{STFT}{Short-Time Fourier Transform}
\newabbreviation{hlb}{HLB}{hyperbaric lifeboat}
\newabbreviation{iruwb}{IR-UWB}{impulse radio ultra-wideband}
\newabbreviation{cw}{CW}{continuous wave}
\newabbreviation{fmcw}{FMCW}{frequency-modulated continuous wave}

\def\BibTeX{{\rm B\kern-.05em{\sc i\kern-.025em b}\kern-.08em
    T\kern-.1667em\lower.7ex\hbox{E}\kern-.125emX}}
\begin{document}

\title{Non-invasive Diver Respiration Rate Monitoring in Hyperbaric Lifeboat Environments using Short-Range Radar
\thanks{This work was supported by the Innovate UK Knowledge Transfer Partnership, grant number KTP10827 and by EPSRC under Grant EP/R513349/1.}
}

\author{\IEEEauthorblockN{
Mikolaj~Czerkawski\IEEEauthorrefmark{1},
Fraser~Stewart\IEEEauthorrefmark{1}\IEEEauthorrefmark{2},
Christos~Ilioudis\IEEEauthorrefmark{1},
Craig~Michie\IEEEauthorrefmark{1},
Ivan~Andonovic\IEEEauthorrefmark{1},
Robert~Atkinson\IEEEauthorrefmark{1},\\
Maurice~Coull\IEEEauthorrefmark{2},
Donald~Sandilands\IEEEauthorrefmark{2},
Gareth~Kerr\IEEEauthorrefmark{2},
Carmine~Clemente\IEEEauthorrefmark{1},
Christos~Tachtatzis\IEEEauthorrefmark{1}
}
                                    % ...

\IEEEauthorblockA{\IEEEauthorrefmark{1}
Department of Electronic and Electrical Engineering, University of Strathclyde, Glasgow, G1~1XW, UK}
\IEEEauthorblockA{\IEEEauthorrefmark{2}
Fathom Systems Ltd.: Badentoy Crescent, Portlethen AB12 4YD, UK}
}

\maketitle

\begin{abstract}
The monitoring of diver health during emergency events is crucial to ensuring the safety of personnel. A non-invasive system continuously providing a measure of the respiration rate of individual divers is exceedingly beneficial in this context. The paper reports on the application of short-range radar to record the respiration rate of divers within hyperbaric lifeboat environments. Results demonstrate that the respiratory motion can be extracted from the radar return signal applying routine signal processing. Further, evidence is provided that the radar-based approach yields a more accurate measure of respiration rate than an audio signal from a headset microphone. The system promotes an improvement in evacuation protocols under critical operational scenarios.
\end{abstract}


\section{Introduction}
\label{sec:introduction}

    The maintenance of offshore infrastructures within the oil and gas sector is critical for uninterrupted production. The current capability of underwater remotely operated vehicles (ROVs) to execute the full range of key maintenance tasks is limited and consequently, the requirement for human intervention remains. The reliance on divers to function at significant underwater depths, in turn, places severe physical and mental demands motivating the obligation to develop protocols and support systems that place the safety of the diver at their core. For extreme depths, the decompression time is significant and can be on the order of days, impacting negatively on the efficiency of the execution of work plans. Saturation diving is a technique for reducing the number of decompressions to one through pressurising divers to the working pressure in living chambers on a dive support vessel. Divers are then transferred to the work site, often hundreds of metres below the surface in a diving bell, a pressurised transfer chamber.
    
    Ensuring diver safety in the extreme hyperbaric environment is an obligation. Divers cannot be removed easily from their operational living spaces in the event of an emergency due to the significant pressure difference between the internal and external chamber environments. Therefore, divers are evacuated in a \gls{hlb}, a rescue vehicle at the same chamber pressure launched from the vessel. 
   
    A maximum of 24 divers can be confined in the restricted space of a rescue vehicle and can spend up to 72 hours in a \gls{hlb} to be transported to a hyperbaric land-based reception facility for decompression. The \gls{hlb} chamber is compact, with a diameter of approximately 1.8m and length as small as 5m~\cite{jfd_website}, and consequently, even mild sea conditions result in challenging conditions inside the chamber, necessitating that divers be strapped into seats for the duration. It is therefore clear that diver safety is critical during \gls{hlb} evacuation.
    
    \begin{figure}[b]
        \centering
        \includegraphics[width=0.9\columnwidth]{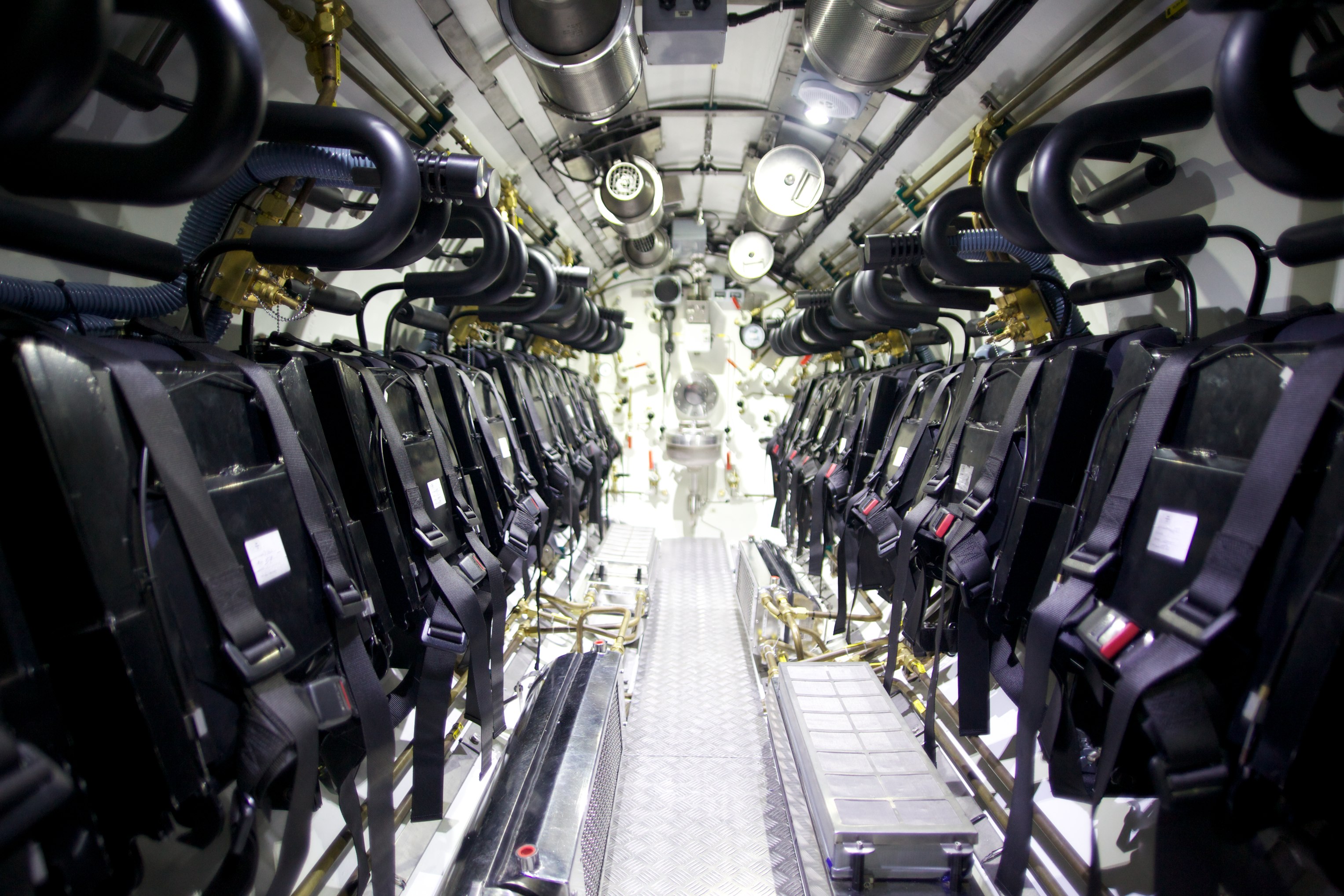}
        \caption{The inside of a hyperbaric lifeboat demonstrating the cramped diver environment}
        \label{fig:inside_0}
    \end{figure}

    An advance that enhances safety is to monitor individual diver health within the \gls{hlb}. Diver health data is then relayed to shore where clinicians can triage their medical condition and prepare the most appropriate treatment in advance of arrival at the hyperbaric reception facility. Currently, no diver health monitoring systems have been reported for application in the \gls{hlb} operational environment. CSMTS is a commercial \gls{hlb} system that predominately monitors chamber environmental conditions and relays that data to shore (CSMTS, Fathom Systems Limited., Portlethen, Aberdeenshire, UK); however currently the system does not provide data on diver health. A closely related system is the Dan Medical D-MAS HyperSat (D-MAS HyperSat, DanMedical Limited, Chipping Norton, Kingham, UK) which measures diver vital signs including heart rate, blood oxygen, blood pressure, and core temperature but is designed for use in living chambers only. Further, D-MAS HyperSat is a reactive solution as it does not continually monitor the divers nor relay data to shore. From a deployment perspective, a significant amount of additional cabling would have to be managed in an already cramped environment within a \gls{hlb}.
            
    \begin{figure}[b]
        \centering
        \includegraphics[width=\columnwidth]{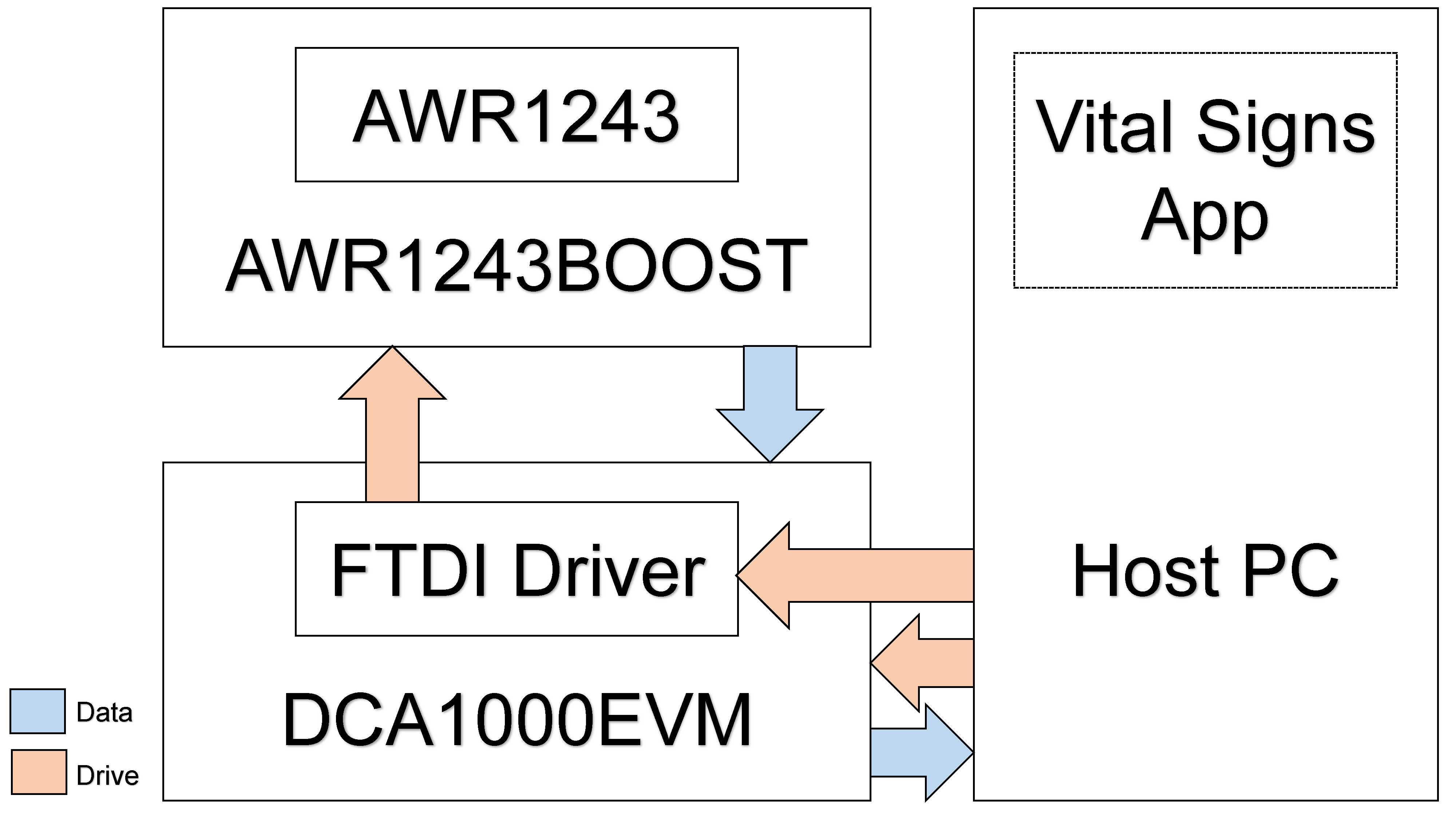}
        \caption{Radar System Diagram}
        \label{fig:diagram}
    \end{figure}
    
    There is a compelling rationale to monitor diver health within \gls{hlb} to inform on the optimum evacuation during emergency events. Any health monitoring solution must have an established route to deployment through integration within existing monitoring platforms such as CSMTS. The features of a fit-for-purpose solution are heavily governed by the already cluttered and confined \gls{hlb} environment. Consequently, health monitoring devices must be wireless reducing the overhead of cabling and implementation complexity within the chamber. Further, the system must not place any significant demands for input by an individual diver as the stress that divers are under during evacuation is significant. Lastly, the system must capture data reliably irrespective of the environmental factors e.g. excessive and unpredictable vehicle motion.  Integrating the health monitoring system into the diver's seat is a potential solution that allows the diver and seat to move together resulting in reduced motion effects.    
    
    The paper reports on the development of a radar-based non-invasive system for the estimation of the respiration rate of divers derived from the motions of the human chest cavity captured through the concomitant Doppler shifts in the reflected signal. There has been a relatively large number of reported applications of radar to measure human vital signs, specifically the respiration rate and the heart rate \cite{Schleicher2013,Yang2019,Hu2016, Lin1975,Li2008,Mostafanezhad2013,Xiong2017,Rahman2018, Mostov2010, Anitori2009,Wang2014,Othman2017,Chang2018,Ahmad2018}. A radar-based system can perform continuous and non-intrusive tracking of the cardiopulmonary activity. Examples of almost all radar modes have been shown to have potential in this application, ranging from \gls{iruwb} Radar\cite{Schleicher2013,Yang2019,Hu2016} through \gls{cw} \cite{Lin1975,Li2008,Mostafanezhad2013,Xiong2017,Rahman2018} to \gls{fmcw}\cite{Mostov2010, Anitori2009,Wang2014,Othman2017,Chang2018,Ahmad2018} Radar. While the detection of vital signs can rely solely on the Doppler information with no need to detect the range, the latter can help to isolate the signal of interest and consequently, filter out clutter and interference.
    
    The paper is organized as follows. Section~\ref{sec:setup} describes the experimental arrangement, including the detail of the characterisation of the static component present in the received signal viz. in an empty chamber. The signal processing steps applied to the measurements are summarized in Section~\ref{sec:processing}. Section~\ref{sec:results} details the results and finally, Section~\ref{sec:conclusion} draws key conclusions arising out of the study.

    \begin{figure}[t]
        \centering
        \includegraphics[width=0.9\columnwidth]{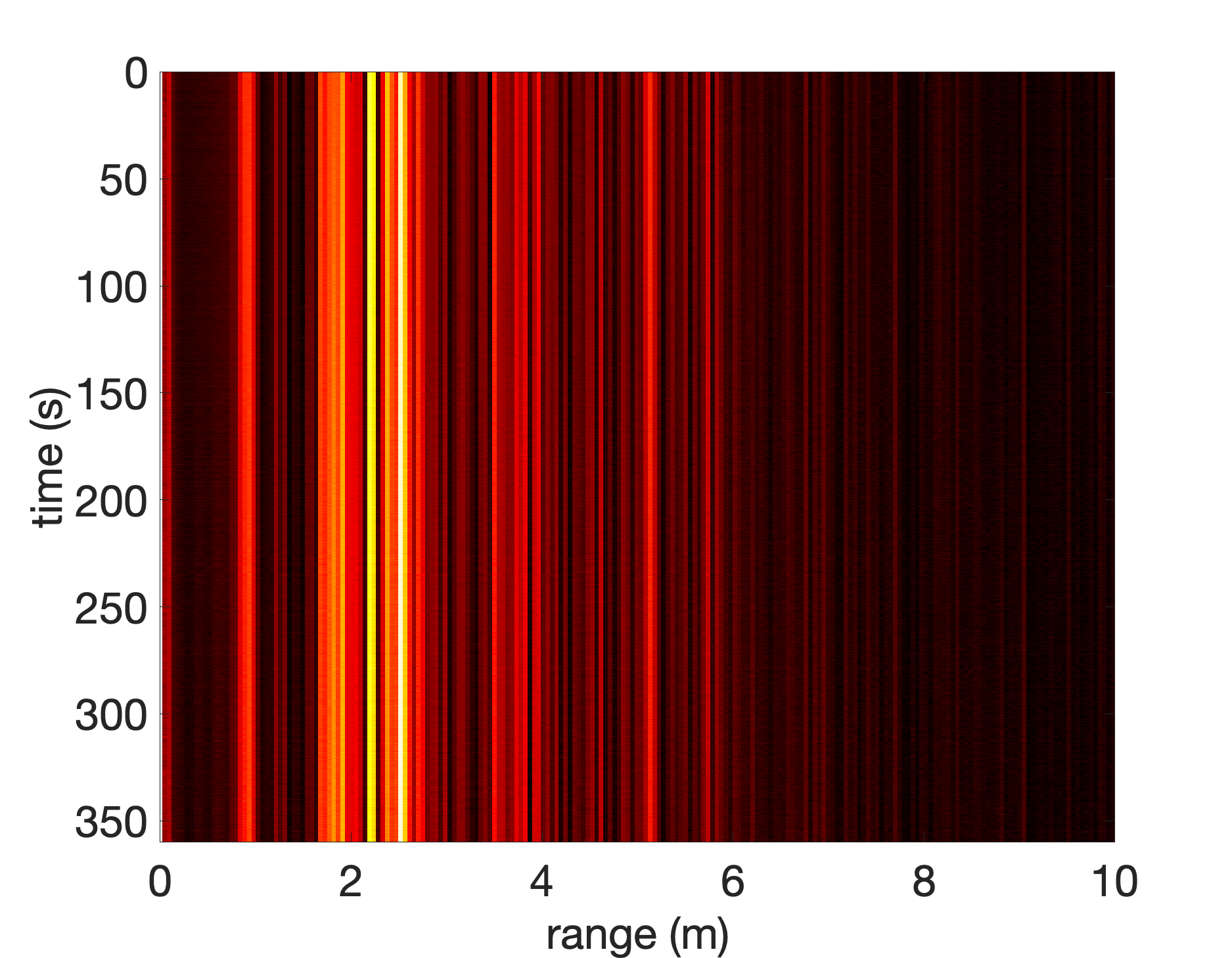}
        \caption{Range time map of the empty chamber}
        \label{fig:empty_range}
    \end{figure}
    
    \begin{figure}[b]
        \centering
        \includegraphics[width=\columnwidth]{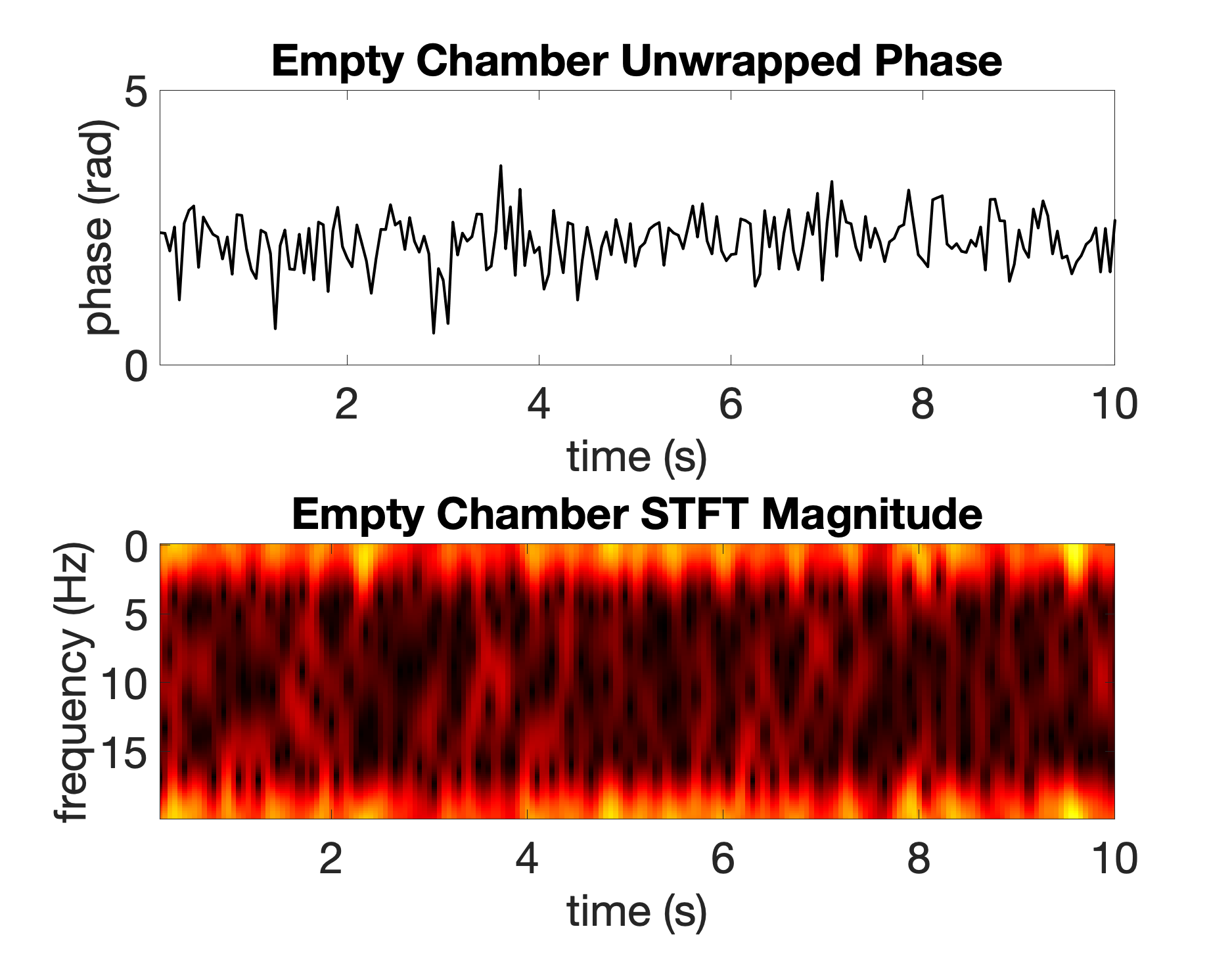}
        \caption{Phase and STFT extracted from the range bin at 1.49 m}
        \label{fig:empty_phase}
    \end{figure}

\section{Experimental Arrangement and Procedure}
\label{sec:setup}

    \begin{figure}[t]
        \centering
        \includegraphics[width=\columnwidth]{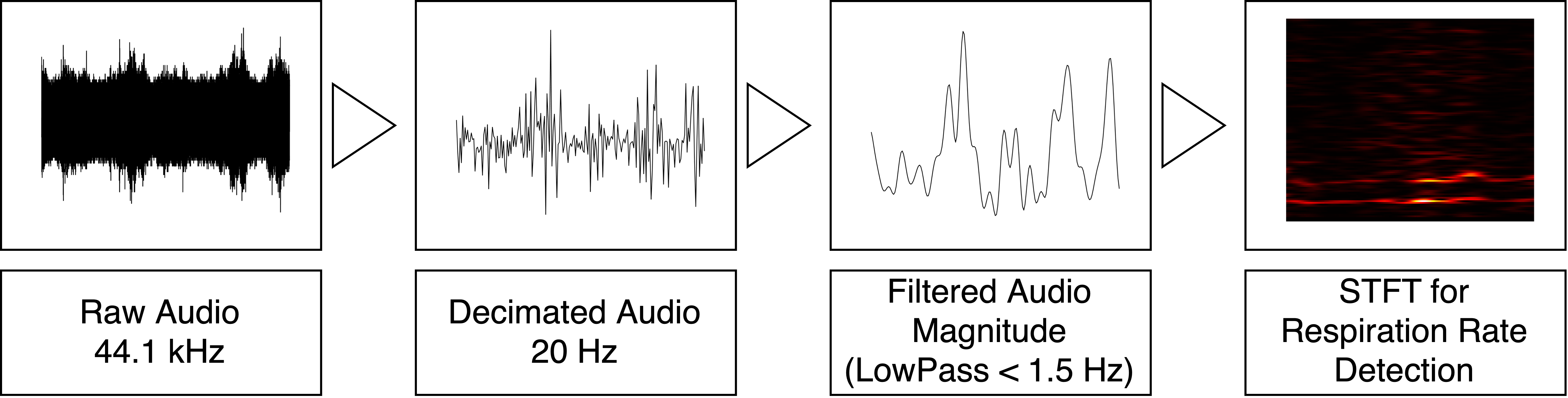}
        \caption{Audio Processing Pipeline}
        \label{fig:audio_processing}
    \end{figure}
    
    \begin{figure}[b]
        \centering
        \includegraphics[width=\columnwidth]{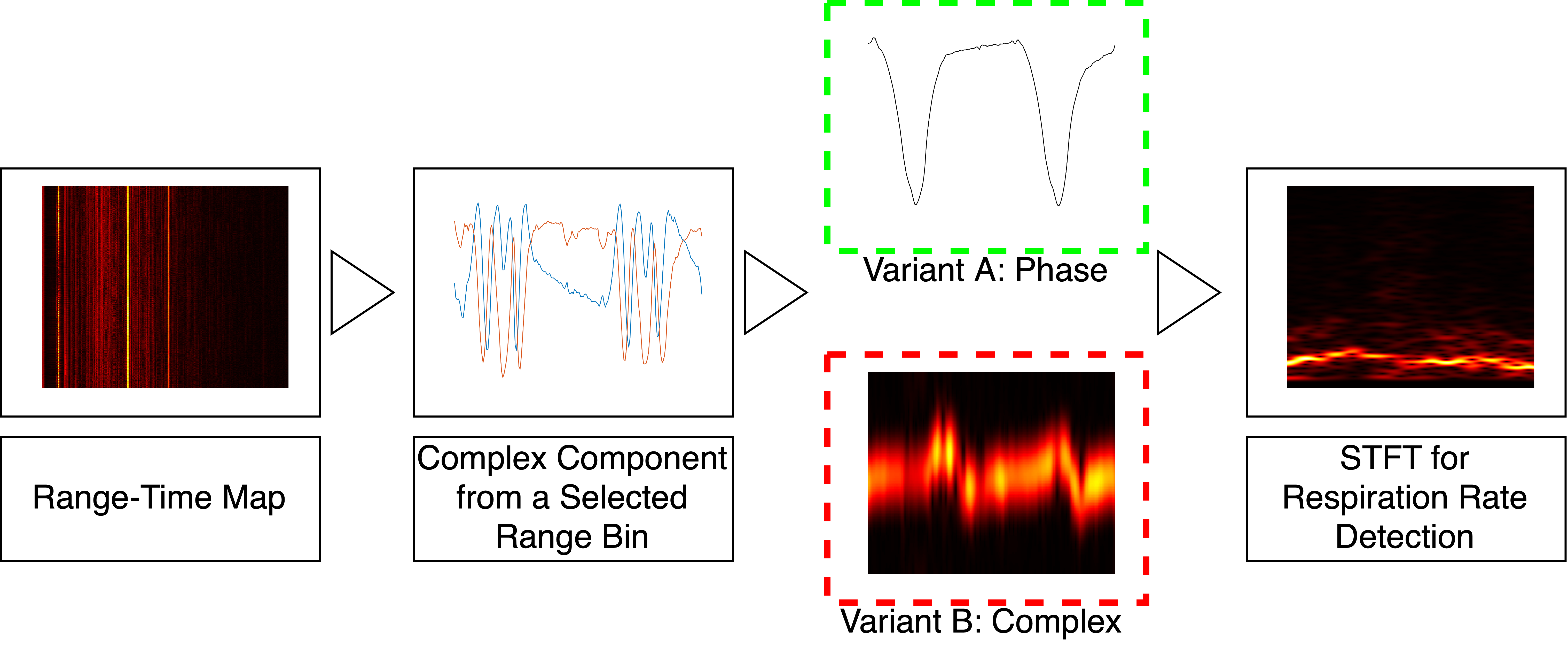}
        \caption{Radar Processing Pipeline}
        \label{fig:radar_processing}
    \end{figure}

    The radar system relies on the AWR1243 module from Texas Instruments, driven via the mmWaveLink library interface, as illustrated in Figure~\ref{fig:diagram}. A custom data acquisition application has been used to drive, trigger, and integrate all elements of the system. The beat signal data is streamed over a high-speed interface to an FPGA-based capture card DCA1000EVM and containerised into datagram packets streamed over a 1 Gbps Ethernet link. The application currently supports capturing the recorded data to a binary file that can subsequently be subject to further analysis.
    
    The truthing data of the target respiration rate was acquired through audio recordings using a headset microphone. The audio stream was triggered at the same time as the radar measurement, coordinated by the acquisition application.

    All data was gathered within a diver lifeboat environment, a constrained space with metal walls; the metal enclosure can contribute to a significant number of multi-path reflections. Additionally, with multiple test targets (divers), the multi-path reflections can interfere with each other in a more complex manner. Figure \ref{fig:inside_0} shows the inside of a hyperbaric lifeboat chamber. The radar device was mounted on a tripod mid-point between the walls, with the antenna aligned to centre of the target's chest. 
    
     A recording of an empty chamber was obtained with the radar antenna placed in the same position as for all other measurements, to investigate the characteristics of the lifeboat environment with no targets present and in order to test the static component of the range map, one of the core elements of the description of the environment. Figure~\ref{fig:empty_range} shows the range-time map computed for the empty chamber recording. The vertical fringes illustrate that the range profile remains approximately constant throughout the 6 minutes of recording. The brightness of the fringes corresponds to the level of reflected power at various ranges. A number of reflections of comparable power between 0 meters and 6 meters is observed in the range profile. The radius of the lifeboat chamber (approximately a meter of magnitude) is significantly less than 6 meters, suggesting that a number of paths are generated by a single reflection at least from a lifeboat wall, unlike the case of direct reflection.
    
    The strongest reflection was selected to investigate the phase component of the signal. Figure \ref{fig:empty_phase} shows a trace of the unwrapped phase (top) and a time-frequency representation of the signal (bottom). The details of the technique to extract these signals of interest are given in the following section, subsequently applied to generate the results presented in Section~\ref{sec:results}.
                       
    \begin{table}[t]
        \centering
        \caption{\gls{stft} Parameters}
        \begin{tabular}{|l | l|}
            \hline
            Window Length & 60 s \\
            Frequency Resolution & 1 bpm \\
            Window Overlap & 59.95 s\\
            Window Shape & Blackman\\
            \hline
        \end{tabular}
        \label{tab:stft_properties}
    \end{table}

\section{Signal Processing}
\label{sec:processing}
    \subsection{Control Signals}
    
        The truthing data was derived from audio recordings sampled at 44.1 kHz captured by a headset microphone. The processing of the audio data stream consisted of a decimation - to match the radar frame rate of 20 Hz (through anti-aliasing with an FIR filter of order 20) - and low-pass filtering of the signal amplitude as illustrated in Figure~\ref{fig:audio_processing}. The low-pass filter was an FIR filter with 60 dB stop-band attenuation and 1.5 Hz pass-band frequency. The resultant reference signal approximates the power of the recorded audio, dominated by inhalation and exhalation sounds (particularly the latter). Finally, in order to extract the respiration rate, a \gls{stft} was applied with the parameters presented in Table~\ref{tab:stft_properties}. The same \gls{stft} parameters were used for the pre-processed radar signal.

        \begin{figure*}[t]
            \centering
            \begin{tabular}{c c c}
            \includegraphics[width=0.32\textwidth]{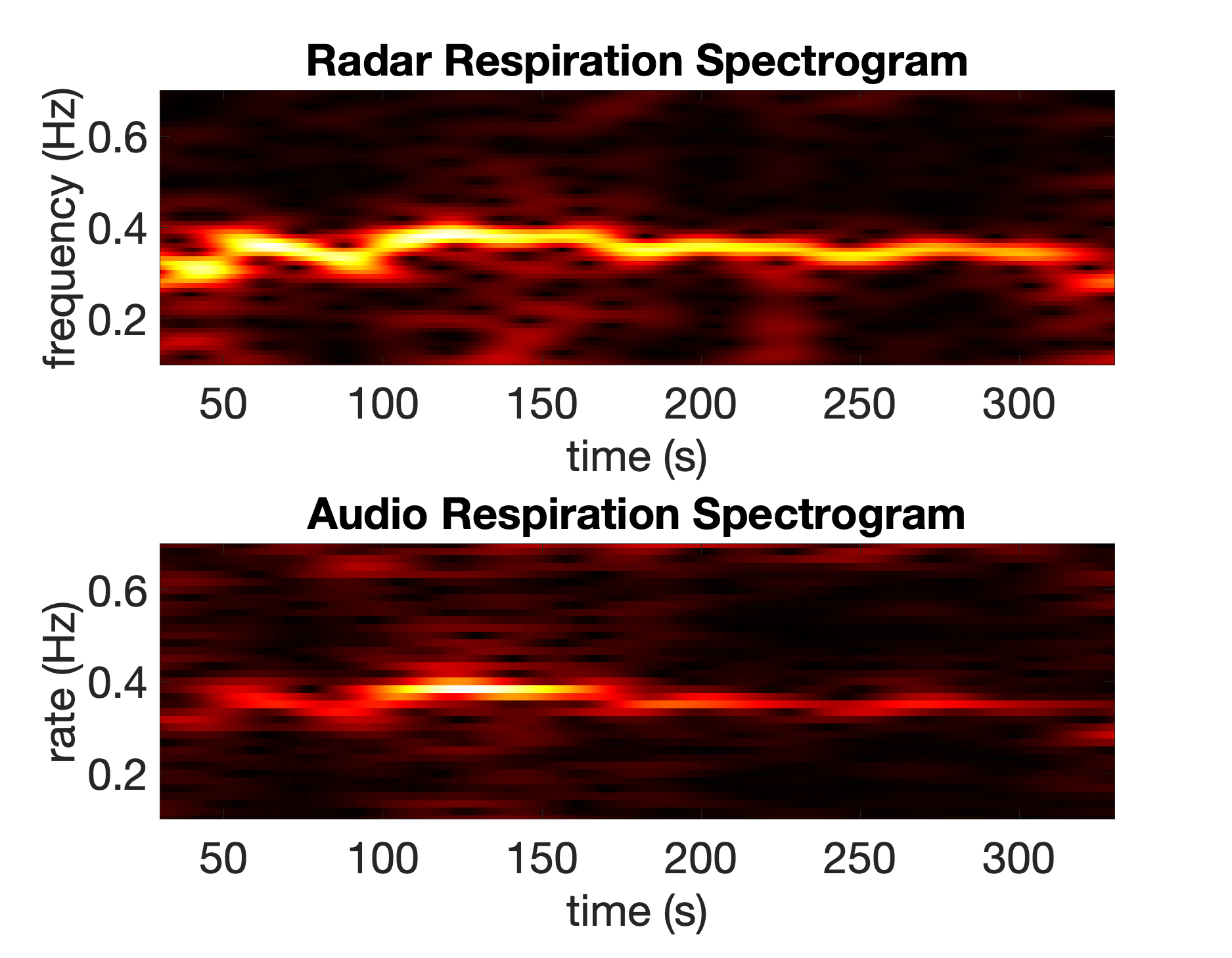} & \includegraphics[width=0.32\textwidth]{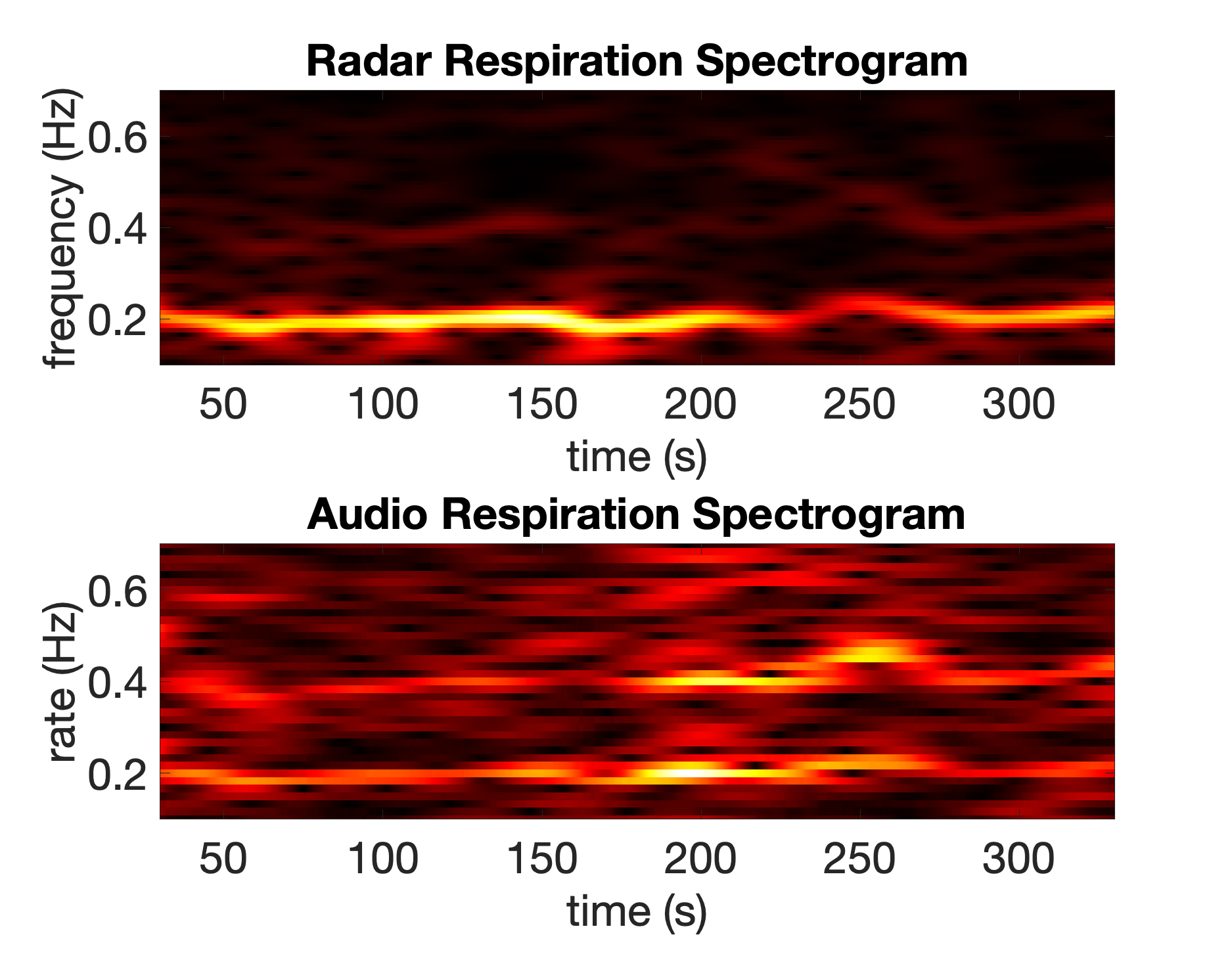} & \includegraphics[width=0.32\textwidth]{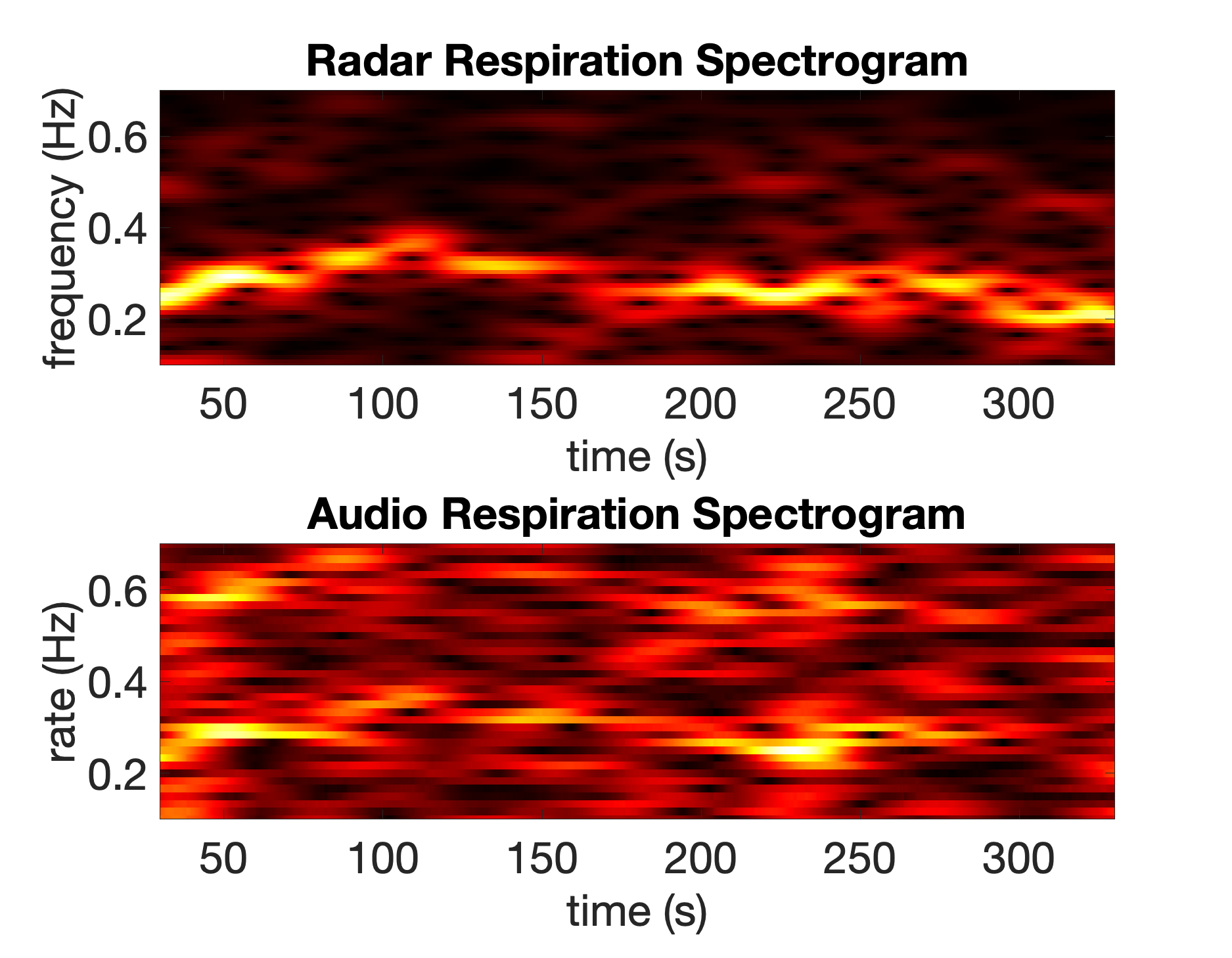}\\
            (a) & (b) & (c)
            \end{tabular}
            
            \caption{Comparison of \gls{stft} output for the test subject 1 (a), 2 (b) and 3 (c)}
            \label{fig:c2}
        \end{figure*}
            
    \subsection{Radar Signal}
    
        \begin{figure}
            \centering
            \begin{tabular}{c}
            \includegraphics[width=0.95\columnwidth]{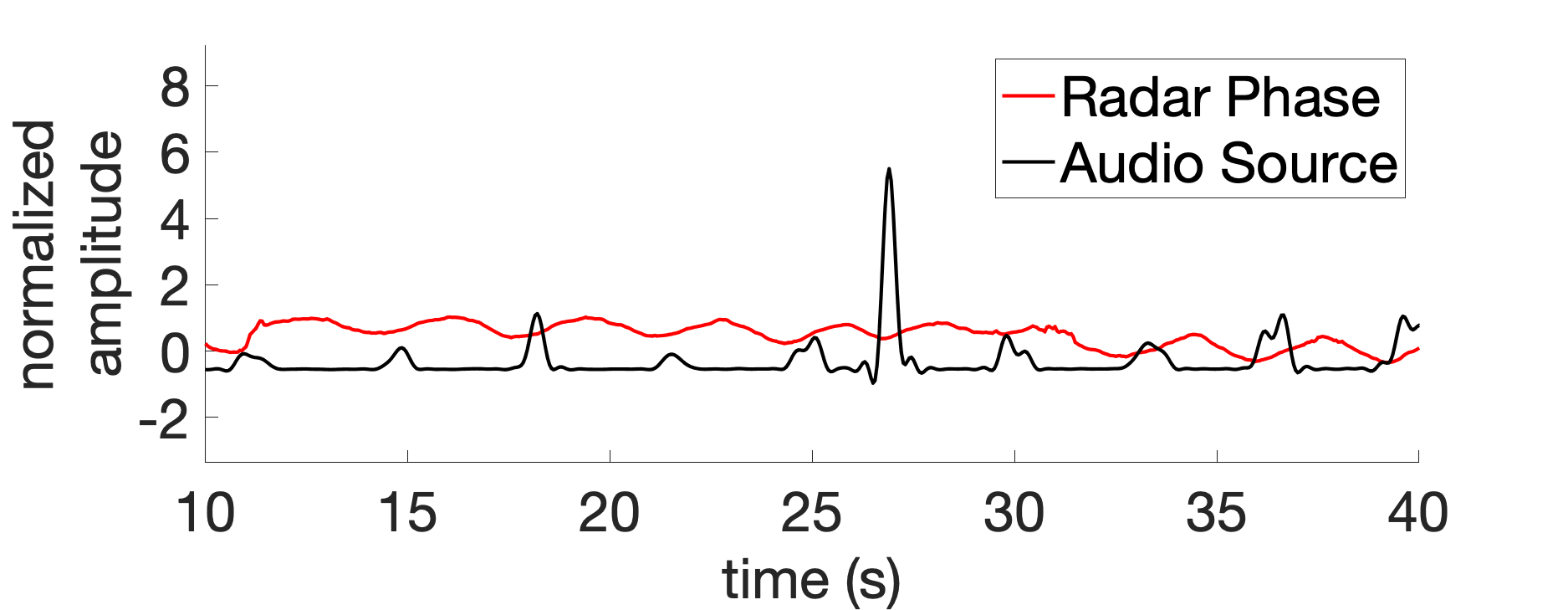} \\
            (a)\\
            \includegraphics[width=0.95\columnwidth]{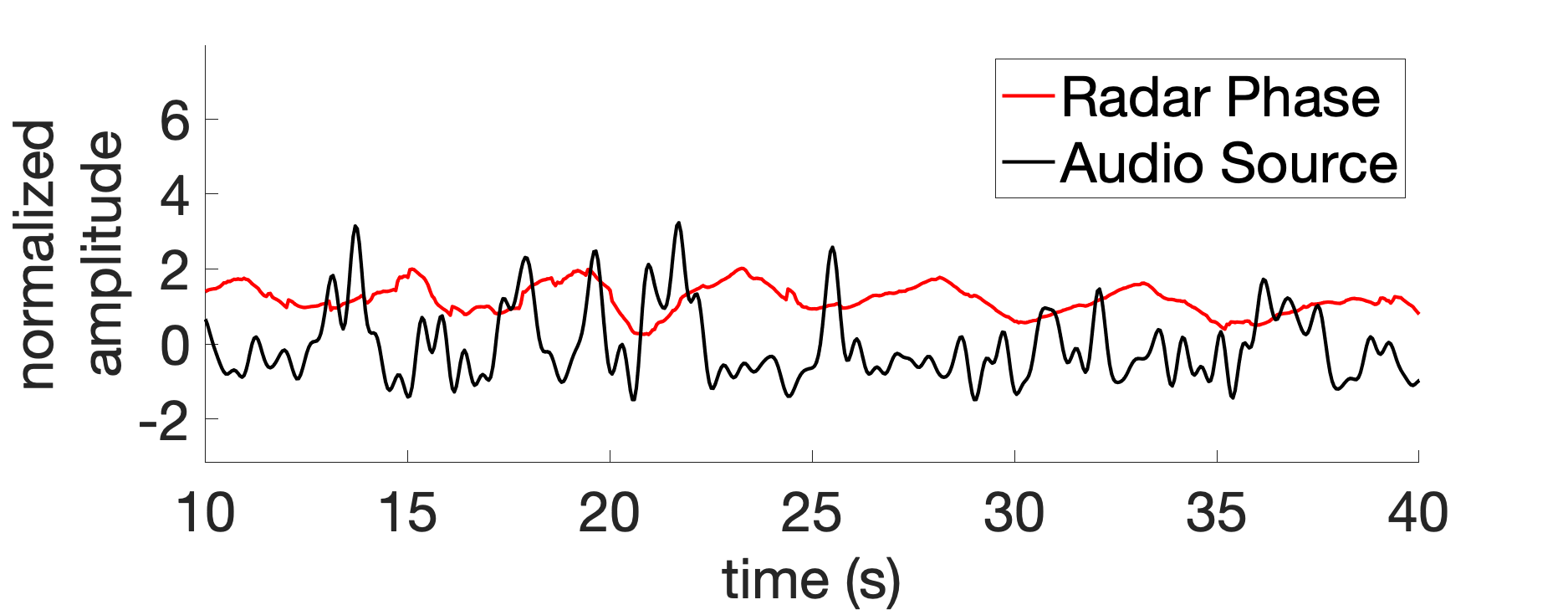}\\
            (b)\\
            \includegraphics[width=0.95\columnwidth]{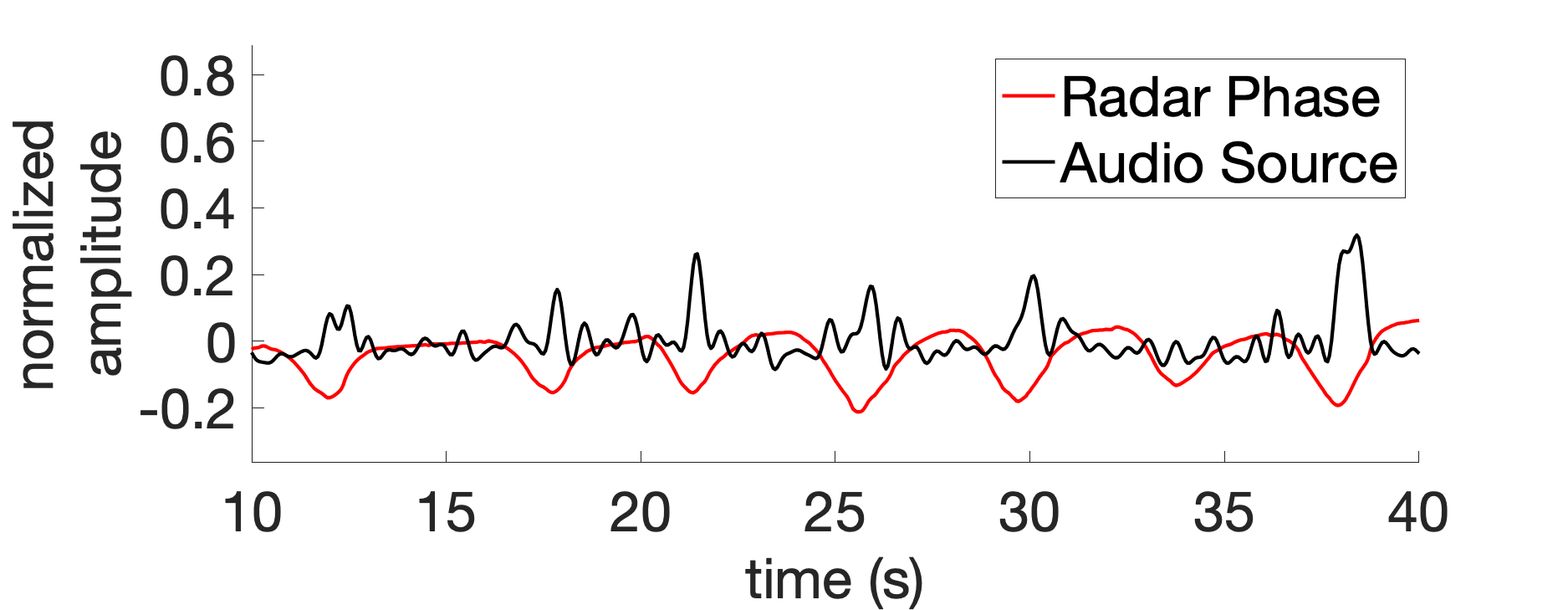}\\
            (c)
            \end{tabular}
            
            \caption{Comparison of normalized radar phase and audio traces for the test subject 1 (a), 2 (b) and 3 (c)}
            \label{fig:c1}
        \end{figure}
    
       A range-time map obtained using a radar in \gls{fmcw} mode is subject to a sequence of processing steps leading to the estimation of the respiration rate, as illustrated in Figure~\ref{fig:radar_processing}. First, the range bin with the strongest reflected power between 10 cm and 80 cm is selected as the target chest is expected to be in that region. The complex reflection signal from the isolated range bin is then subject to de-trending, phase extraction and unwrapping (illustrated as Variant A in Figure~\ref{fig:radar_processing}). The unwrapped signal is finally subject to \gls{stft} with the parameters summarized in Table~\ref{tab:stft_properties}.\footnote{It may be worth noting that the phase extraction is not the only way of obtaining a temporal trace useful for Doppler processing. In fact, some works\cite{Czerkawski2020} suggest that this approach may be inappropriate when multiple independent motion components are to be isolated and suggest the application of time-frequency processing to the complex reflection signal instead (Variant B in Figure~\ref{fig:radar_processing}). However, in this study, since only the dominating respiratory trace is of interest justifies the simplified processing approach that was applied.}

\section{Results}
\label{sec:results}

    A set of three recordings was been obtained; one for each test subject. The baseline scenario involves a subject remaining relatively still and breathing freely for 6 minutes.

     The temporal traces corresponding to the respiratory activity from both the audio and the radar signal as shown in Figure~\ref{fig:c1} were extracted Following the approach described in Section~\ref{sec:processing}. In the case of the latter, the phase-extraction variant is applied. The radar phase trace appears to be cleaner (in the sense that it contains less high-frequency components) and displays a degree of resemblance to traces characteristic of chest motion. The audio trace appears fairly clean for test subject 1, however, it only registers the impulsive events of exhalation. For test subjects 2 and 3, the audio trace appears relatively noisy and contains higher frequencies than the radar phase trace. Notwithstanding the noisy characteristic of the audio signal, it is important to note that subjects were making both inhalation and exhalation sounds of comparable power; thus the dominant frequency is double of the respiration rate (this phenomenon has been observed for test subject 2). Interestingly, it appears that even at this initial inspection stage, the control audio signal suffers from higher distortion and interference than the radar signal.
    
    The radar and audio traces from Figure~\ref{fig:c1} are subject to further time-frequency analysis in order to extract the periodic components dominant within the signals; here, \gls{stft} is applied with the parameters displayed in Table~\ref{tab:stft_properties}. The magnitude of the resultant time-frequency distributions is shown in Figure~\ref{fig:c2}, further proof that the spectrum derived from the radar is of higher sparsity than that of the audio signal. Moreover, it is also evident that the audio-based \gls{stft} contains significantly more noise and strong harmonic components for test subjects 2 and 3.
    
\begin{figure}
        \centering
        \begin{tabular}{c}
        \includegraphics[width=\columnwidth]{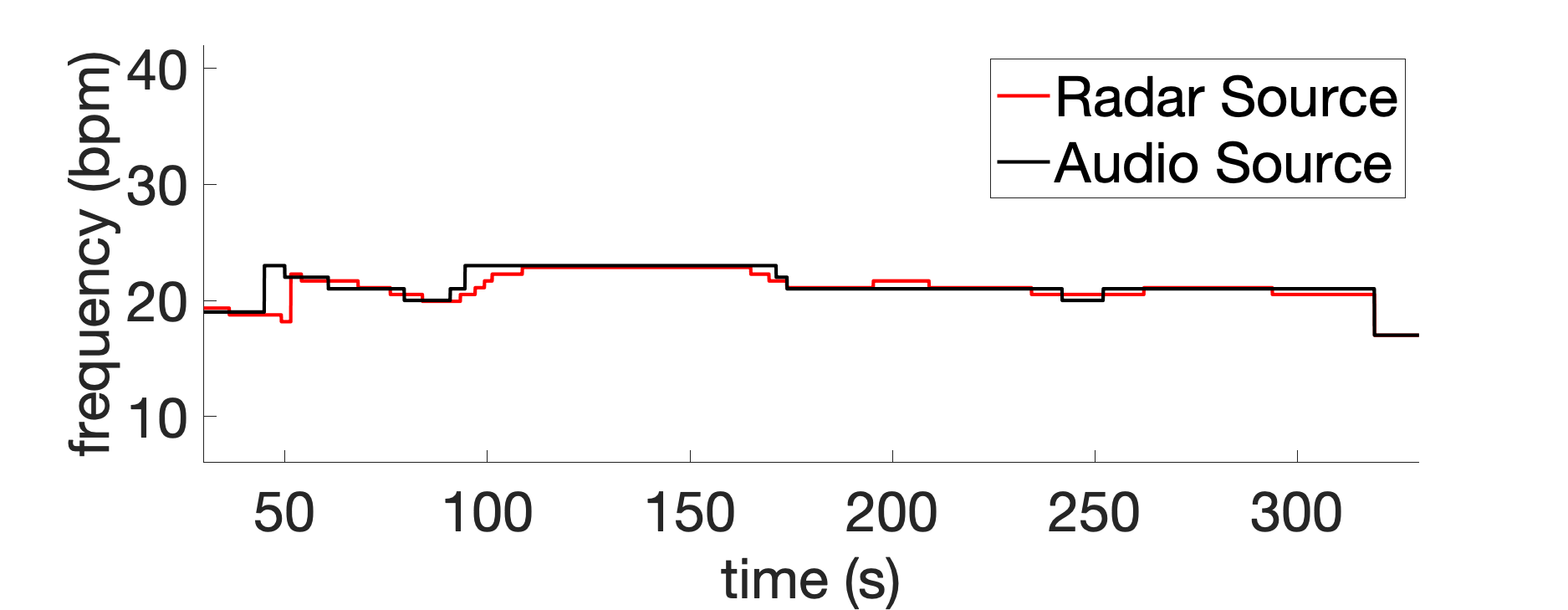}\\
        (a)\\
        \includegraphics[width=\columnwidth]{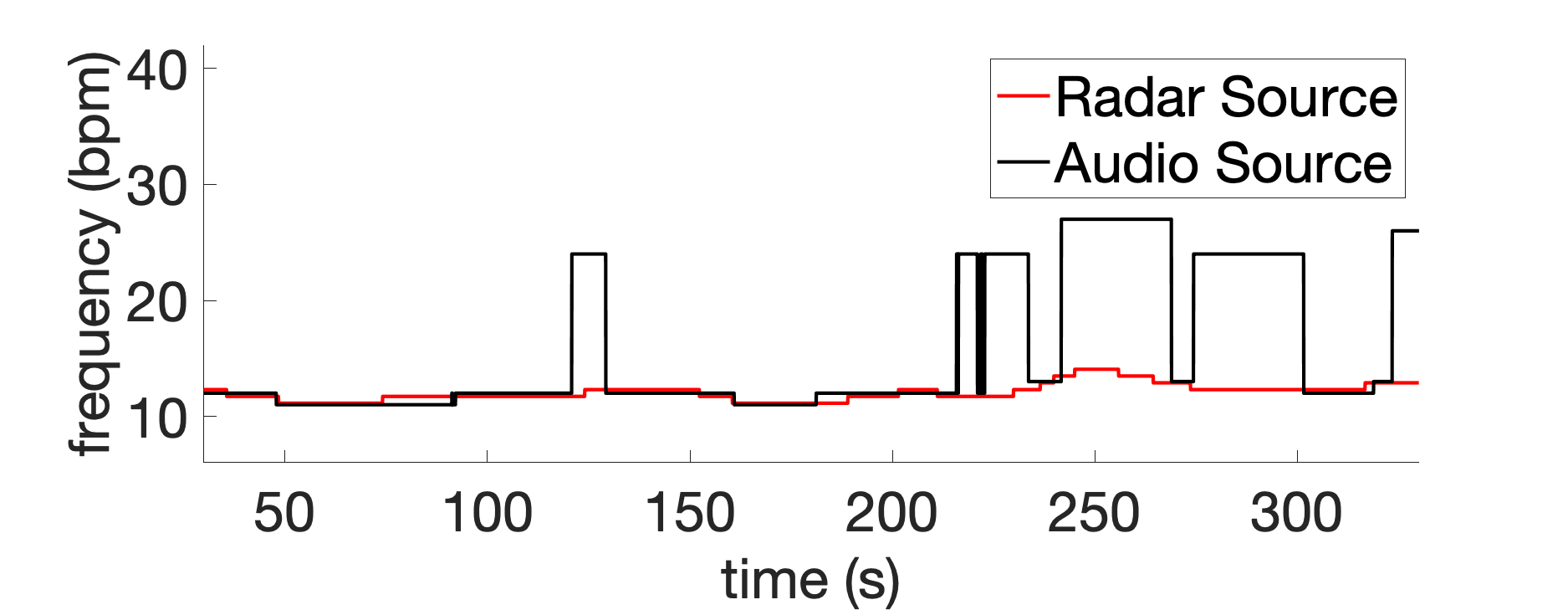}\\
        (b)\\
        \includegraphics[width=\columnwidth]{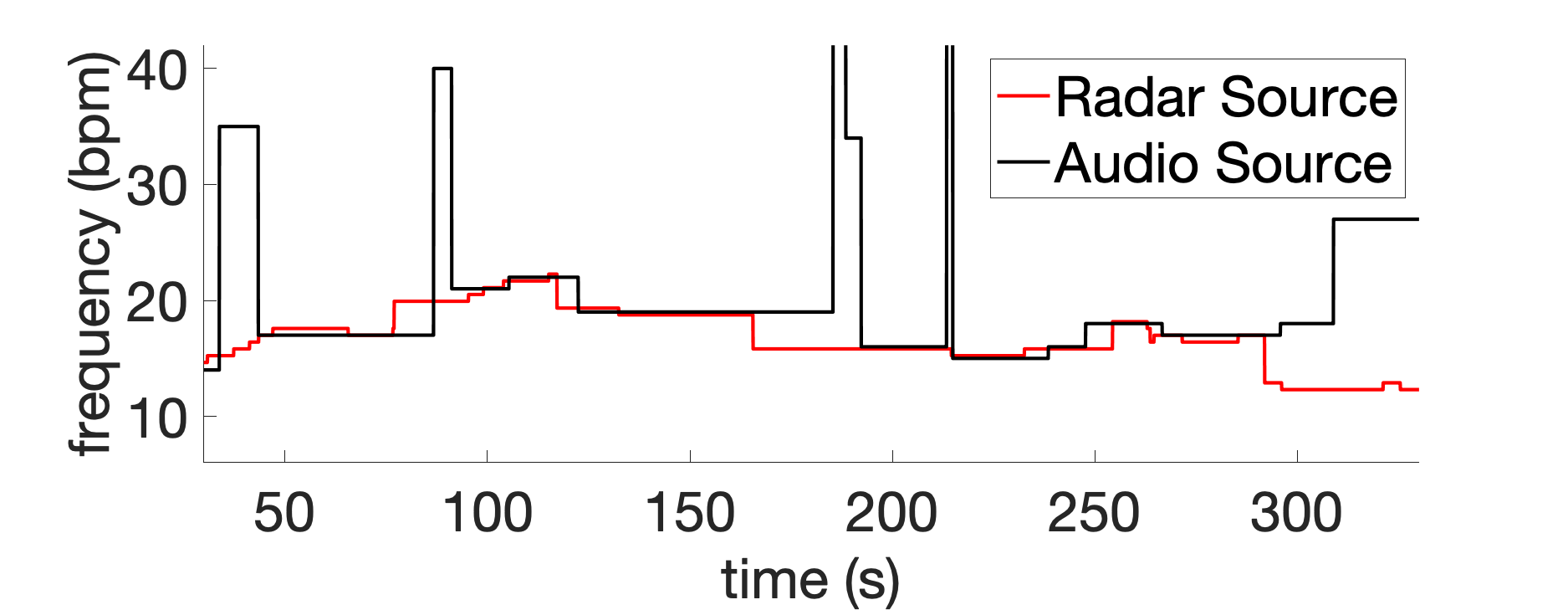}\\
        (c)
        \end{tabular}
        
        \caption{Comparisons of extracted respiration rate estimations for the test subject 1 (a), 2 (b) and 3 (c)}
        \label{fig:c3}
    \end{figure}

    Computing the frequency through identifying the highest magnitude for each time instant was the direct approach utilised to extract the respiration rate from a \gls{stft}; the results are displayed in Figure~\ref{fig:c3}. The strong harmonics and noise in the audio signal distorted the estimation of the  frequency, resulting in random and erratic variations of the extracted rate. On the other hand, the results derived from the radar-based \gls{stft} were significantly more consistent.

    It is evident that the two techniques investigated in the study yield differences in the performance quality of the estimation of the respiratory rate. The relatively straightforward signal processing approach described in Section~\ref{sec:processing} was applied as the key goal was to provide evidence that a radar technique is a viable approach to remote respiration rate monitoring within challenging environments. In addition, initial proof is provided that a radar-based solution is easier to post-process than audio implementations.

\section{Conclusions}
\label{sec:conclusion}
    Results are provided that prove the applicability of a radar-based respiration rate monitoring system in deep sea diver rescue vehicle environments. The system is capable of the continuous monitoring of human respiration rate in a challenging and constrained environment of a \gls{hlb}. Moreover, the performance of the estimation is not significantly impaired by multi-path reflections for a single target scenario in \gls{fmcw} radar mode.
    
    There are a few significant advantages of the proposed system. Most importantly, it allows for remote, non-intrusive monitoring of the target's respiration rate, which is immensely important in emergency scenarios when a \gls{hlb} is utilised. The respiration rate is monitored continuously and requires no attention of the monitored target as long as the target remains relatively stationary. Furthermore, it is discovered that an \gls{fmcw} radar signal can potentially be a source of cleaner respiration trace than the audio recorded by a headset microphone. Lastly, it should be stressed that there may be a lot of other information that can be extracted from the same radar signal. Apart from the respiratory activity, there is potential to monitor the heart rate in the same way or to observe target motion and classify motion events. Understandably, this information can lead to significant advances to the evacuation practices and ultimately, improvement of diver's safety.
    
    Several directions are to be taken in future work. The experiments carried out for this work involve single target scenarios. Consequently, a set of multi-target experiments could be a valuable development. Similarly, it would be beneficial to test this approach with the test subjects wearing diving equipment and recreate the evacuation scenario more faithfully in order to identify the factual utility and performance of the proposed system. Another interesting direction to take is to attempt to extract more data from the radar signal, such as the heart rate or target's activity.
    
\bibliographystyle{IEEEtran}
\bibliography{IEEEabrv,references.bib}
    
\end{document}